\documentclass[vecphys]{svmult}

\usepackage{makeidx}         
\usepackage{graphicx}        
\usepackage{multicol}        
\usepackage[bottom]{footmisc}
\usepackage{amssymb}
\usepackage{amsfonts}
\usepackage{amsmath}
\def\PKS#1{\,\hbox{$\looparrowleft\!\!\!\!\underline{\,\,\,\,^{#1}\,\,\,\,}\!\!\!\!\looparrowright$}\,}

\makeindex             

\begin{document}

\title*{Funnels in Energy Landscapes}
\author{Konstantin Klemm\inst{1}\and
  Christoph Flamm\inst{2}\and
  Peter F.\ Stadler\inst{1,2,3}}
\institute{%
 $^1$Bioinformatics Group, Department of Computer Science,
        University of Leipzig\newline
 $^2$Institute for Theoretical Chemistry, University of Vienna,
        Austria\newline	
 $^3$Santa Fe Institute, Santa Fe, New Mexico, USA\newline
\texttt{\{klemm,stadler,xtof\}@bioinf.uni-leipzig.de}
}

\maketitle

\begin{abstract}
Local minima and the saddle points separating them in the energy landscape
are known to dominate the dynamics of biopolymer folding. Here we
introduce a notion of a ``folding funnel'' that is concisely defined in
terms of energy minima and saddle points, while at the same time conforming
to a notion of a ``folding funnel'' as it is discussed in the protein folding
literature.
\end{abstract}

\section{Introduction}

The dynamics of structure formation (``folding'') of biopolymers, both
protein and nucleic acids, can be understood in terms of their \emph{energy
landscapes}. Formally, a landscape is determined by a set $X$ of
conformations or states, a neighborhood structure of $X$ that encodes which
conformations can be reached from which other ones, and an energy function
$E:X\to\mathbb{R}$ which assigns the folding energy to each state. In
the case of nucleic acids it has been demonstrated that dynamics features
of the folding process can be derived at least in a good approximation by
replacing the full landscape by the collection of local minima and their
connecting saddle points \cite{Wolfinger:04a}. 

The notion of a \emph{``folding funnel''} has a long history in the protein
folding literature \cite{Bryngelson:87,Leopold:92,Onuchic:97,Galzitskaya:99,
Onuchic:00}.
It arose from the observation that the folding process of naturally evolved
proteins very often follows simple empirical rules that seem to bypass the
complexity of the vast network of elementary steps that is required in
general to describe the folding process on rugged energy
landscapes. Traditionally, the funnel is depicted as a relation of folding
energy and ``conformational entropy'', alluding to the effect that the
energy decreases, on average, as structures are formed that are more
and more similar to the native structure of a natural protein
\cite{Onuchic:04}. It may come as a surprise therefore, that despite the
great conceptual impact of the notion of a folding funnel in protein
folding research, there does not seem to be a clear mathematical definition
of ``funnel''. Intuitively, one would expect that a funnel should be defined in
terms of the basins and barriers of the fitness landscape (since, as
mentioned above, these coarse-grained topological features determine the
folding dynamics). Furthermore, it should imply the ``funneling'' of
folding trajectories towards the ground state of the molecule.

\section{Folding Dynamics as a Markov Chain}

We consider here only finite discrete conformations spaces $X$ with a
prescribed set of elementary moves of transitions that inter-convert
conformations. In the following we write $M(x)$ for the set of conformations
accessible from $x\in X$. For example, $X=\{-1,+1\}^N$ in spin-glass
setting, where flipping single spins is the natural definition of a
move. In the case of RNA or protein folding, the breaking and formation of
individual contacts between nucleotides or amino acids, resp., is the most
natural type of move set \cite{Flamm:00a}.

The dynamics on $X$ is modeled as usual by the 1st order Markov chain
with  Metropolis transition probabilities 
\begin{equation}
  \begin{split}
p(y|x) & = \frac{1}{|M(x)|} \min \{1, \exp (- \beta (E_y - E_x))\} 
\qquad\textrm{for } y \in M(x)\\
p(x|x) & = 1 - \sum_{y\in M(x)} p(y|x)
  \end{split}
\end{equation}
All other transition probabilities are zero.

We will be interested in the average time $\tau_x$ the system takes to
reach a pre-defined target state $x_0\in X$ when starting at state $x\in
X$, given by the recurrence
\begin{equation}
\tau_x =  \sum_{y\in M(x)} p(y|x) \tau_{y} + p(x|x) \tau_{x} + 1
\end{equation}
with $\tau_0 = 0$ (target state).

\begin{figure}[t]
\centering
\includegraphics[width=\textwidth,clip=]{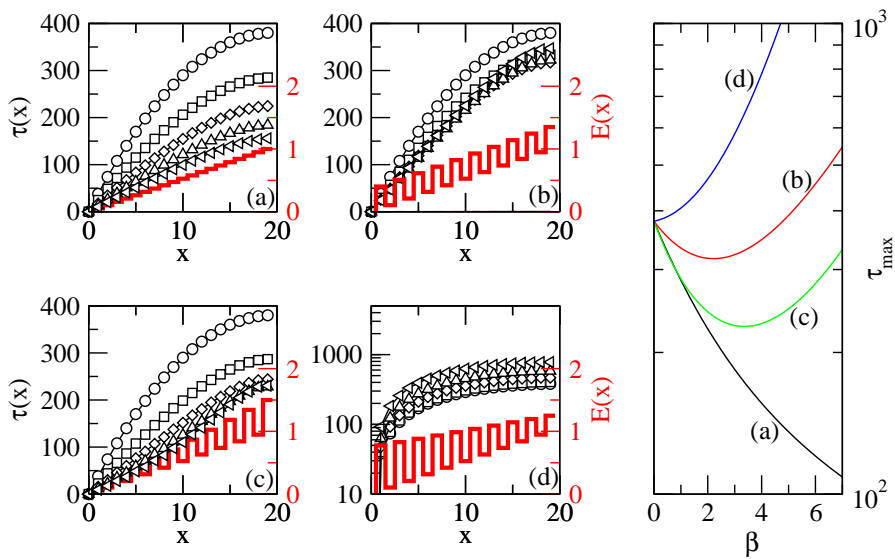}
\caption{(a-d) Dynamics on one-dimensional
energy landscapes $E(x)$ (thick curves).  Thin curves show the average
first passage time for state $x_0=0$ when starting from given state $x$,
for temperatures $\beta=0$ (circle), $\beta=1$ (square), $\beta=2$
(diamond), $\beta=3$ (triangle up), $\beta=4$ (triangle left).  R.h.s.\
panel: temperature dependence of first passage times in the landscapes
(a-d); $\tau_{\rm max}$ is the average time to reach $x_0$ for the
first time starting at the ``rightmost'' state $x=19$. Slight changes
in the slopes or other details of the landscapes $E(x)$ do not change
the qualitative behavior of $\tau_{\rm max}$ as long as the ordering of
barrier heights is conserved.
}
\label{fig:oned} 
\end{figure}

In order to investigate the physical basis of the ``funneling effect''
we start with a simple 1-dimensional toy model with landscapes
defined over the integers $\{0,\dots\,n\}$, see Fig.\ \ref{fig:oned}.
The time $\tau$ to target crucially depends on the ordering
of barriers. The time to target is shortest when barriers are
decreasing towards the ground state as in panel (c) of Fig.\
\ref{fig:oned}.  The property of decreasing barriers towards the
ground state matches the intuition of folding funnels. In the following
section we generalize it to arbitrary landscapes. 

\section{Geometric Funnels}

A conformation $x\in X$ is a local minimum if $E_x\le E_y$ for all $y\in
M(x)$. Allowing equality is a mere mathematical convenience
\cite{Flamm:02a}. Let $\mathbb{P}_{xy}$ be the set of all walks from $x$ to
$y$. We say that $x$ and $y$ are \emph{mutually accessible at level
$\eta$}, in symbols
  \begin{equation}
    x \PKS{\eta} y\,,
  \end{equation}
if there is walk $\mathbf{p}\in\mathbb{P}_{xy}$ such that $E_z\le\eta$ for
all $z\in\mathbf{p}$, respectively.  The \emph{saddle height} $\hat f(x,y)$
between two configurations $x,y\in X$ is the minimum height at which they
are accessible from each other, i.e.,
\begin{equation} \hat f(x,y) =
\min_{\mathbf{p}\in\mathbb{P}_{xy}} \max_{z\in\mathbf{p}} E_z = \min
\{\eta | x \PKS{\eta} y \}
\end{equation}
The saddles between $x$ and $y$ are exactly the maximal points along the
minimal paths in the equation above. We say, furthermore, that a saddle
point $s$ \emph{directly connects} the local minima $x$ and $y$, if there
are paths $\mathbf{p}_{sx}$ and $\mathbf{p}_{sy}$ from $s$ to $x$ and $y$,
respectively, such that  $E$ is monotonically decreasing along both
$\mathbf{p}_{sx}$ and $\mathbf{p}_{sy}$.

\begin{figure}[t]
  \begin{center}
    \includegraphics[width=\textwidth]{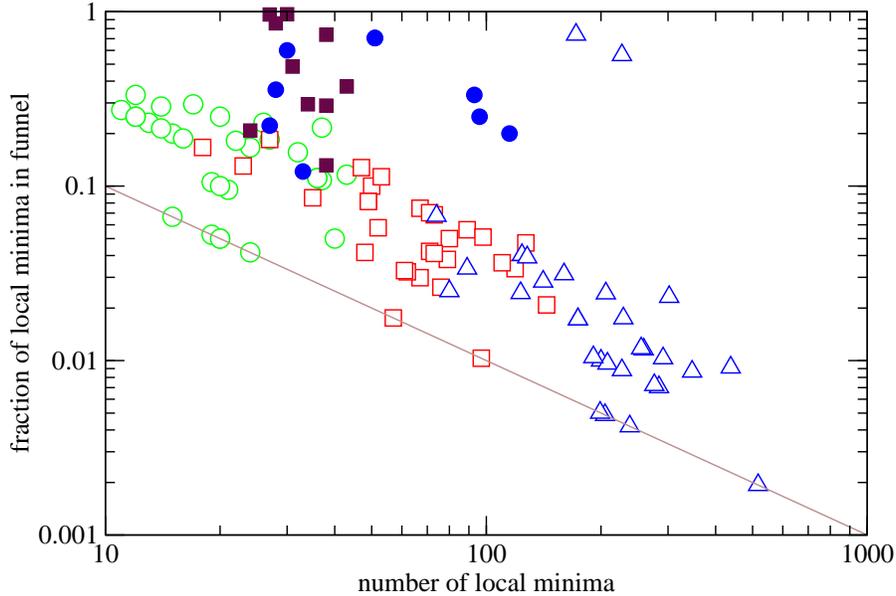}
  \end{center}
  \caption{Fraction of minima belonging to the funnel vs.\ the total number
    of minima found in the landscape. RNA hairpins (filled squares) and
    RNAs with two different near-ground state structures (filled circles).
    The straight line has slope -1. Landscapes with only the ground state
    in the funnel fall on this line. Open symbols are the results for the
    number partitioning problem with (NPP) with sizes $n=8$ (circles), $n=10$
    (squares) and $n=12$ (triangles). For each system size, 30 instances were
    generated by drawing random numbers $a_1,a_2,\dots,a_n$ and $c$
    independently from the unit interval. The energy for a state
    $(x_1,\dots,x^N) \in \{-1,1\}$ is
    $E(x) = |\sum_{i=1}^n x_i a_i + c|$.
    The $c$ represents an extra ``clamped'' degree of freedom to break the
    symmetry under reversal of all spins. This ensures that almost all
    instances have a unique ground state. }
  \label{fig:funnel}
\end{figure}

For simplicity we assume the following weak non-degeneracy
condition for our energy landscape: For every local minimum $x$ 
there is a unique saddle point $s_x$ of minimal height $h(x)=\min_z
\hat f(x,z)$. Note that $s_x$ is necessarily a direct saddle between $x$
and some other local minimum $z$, which for simplicity we again assume to
be uniquely determined. This condition is stronger than local
non-degeneracy but weaker than global non-degeneracy in the sense of
\cite{Flamm:02a}.  In the degenerate case, we consider the set of all
direct saddles of minimum height and the set of the local minima directly
connected to them.

Now we can define \emph{the funnel} of a landscape recursively as the
following set $F$ of states:
\begin{itemize}\setlength{\itemsep}{0pt}
\item[(1)] The ground state is contained in the funnel $F$.
\item[(2)] The local minimum $x$ belongs to the funnel $F$ if a minimum
  saddle $s_x$ connects directly to local minimum in the funnel $F$.
\item[(3)] A state $z$ belongs to the funnel if it is connected by a
  gradient descent path to a local minimum in $F$.
\end{itemize}

Using the above definition, we can recursively partition the landscape into
``local funnels'': Simply remove $F$ from $X$ and recompute the funnel of
the residual landscape.

\begin{figure}[t]
  \begin{center}
    \includegraphics[width=0.65\textwidth]{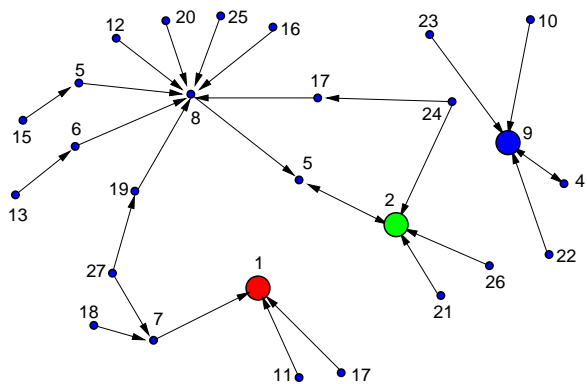}
  \end{center}
\caption{Funnel partitioning for the folding landscape of the RNA sequence
\textsf{xbix} (\texttt{CUGCGGCUUUGGCUCUAGCC}). The landscape falls into
three funnels. In \cite{Wolfinger:04a} it was shown that a large part of
the folding trajectories reach the metastable state 2 whose energy lies 0.8
kcal above the energy of the ground state 1.  }
\label{fig:xbix}
\end{figure}

As one example of biopolymers we consider small artificial RNA sequences
which have been designed either to fold into a single stable hairpin
structure or to have two near-ground state structures that have very few
base pairs in common. In the first case we expect landscapes dominated by
funnels because the \texttt{RNAinverse} algorithm \cite{Hofacker:94a} tends
to produce robustly folding sequences. In the second case we used the
design procedure outlined in \cite{Flamm:00b} to produce sequences that
have decoy structures with moderate to large basins of attraction. The
sequences we use here have a length of 30nt or less, shorter than most
structured RNAs of biological importance.

The \texttt{barriers} algorithm, which efficiently computes local minima
and their separating saddle points from an energy-sorted list of states
\cite{Flamm:00a} can be modified to compute the funnel-partitioning of the
energy landscapes. We will report on this topic elsewhere.

Figure \ref{fig:funnel} shows the fraction of local minima contained in the
funnels of several landscapes. The RNA folding landscapes have folding
funnels comprising a large part of the landscape. The landscapes of RNA
sequences forming hairpins have the largest funnels. In comparison we
plot the relative sizes of funnels for number partitioning problems
of different sizes, as defined in the caption of Fig.\ \ref{fig:funnel}.
These artificial landscapes have significantly smaller funnels than the RNA
folding landscapes. Thus the latter have folding funnels much
larger than expected for random rugged landscapes. Through these large
funnels the folding polymer may be ``guided'' towards the native state.

Figure \ref{fig:xbix} shows an example of an RNA sequence with a strong
kinetic trap studied in detail in \cite{Wolfinger:04a}. In this landscape,
a suboptimal structure has local funnel that covers most of the landscape,
while the ground state is separated by comparably high barriers from almost
all other local optima.

In summary, we have introduced here a rigorous definition of a folding
funnel that is tractable computationally for arbitrary energy landscapes.
In the case of RNA, where the lower fraction of the landscape can be
generated without the need for exhaustively enumerating all configurations
\cite{Wuchty:99}, funnels can be computed explicitly even for sequences
that are of immediate biological interest. Our first computational results
show that the energy landscapes of RNAs typically differ from the rugged
landscapes of spinglass-style combinatorial optimization problems by
exhibiting significantly larger funnels for the ground state. It remains to
be investigated in future work whether this is also true e.g.\ for lattice
protein models. A second important topic of ongoing research is the
question which and to what extent evolutionary processes select molecules
with funnel-like landscapes.

\par\noindent\textbf{Acknowledgments.}
This work was supported in part by the EMBIO project in FP-6
(\texttt{http://www-embio.ch.cam.ac.uk/}).

\bibliographystyle{splncs.bst}
\bibliography{funnel}
\end{document}